\documentclass[%twocolumn,
showpacs,preprintnumbers,amsmath,amssymb]{revtex4}
\usepackage{amsmath}

\begin{document}
%\draft

\tolerance=5000

\def\pp{{\, \mid \hskip -1.5mm =}}
\def\cL{{\cal L}}
\def\be{\begin{equation}}
\def\ee{\end{equation}}
\def\bea{\begin{eqnarray}}
\def\eea{\end{eqnarray}}
\def\beq{\begin{eqnarray}}
\def\eeq{\end{eqnarray}}
\def\tr{{\rm tr}\, }
\def\nn{\nonumber \\}
\def\e{{\rm e}}

\preprint{YITP-07-58}

\title{Modified $f(R)$ gravity unifying $R^m$ inflation with $\Lambda$CDM epoch}
\author{Shin'ichi Nojiri}
\email{nojiri@phys.nagoya-u.ac.jp}
\affiliation{Department of Physics, Nagoya University, Nagoya 464-8602. Japan}
\author{Sergei D. Odintsov\footnote{also at Lab. Fundam. Study, Tomsk State
Pedagogical University, Tomsk}}
\email{odintsov@ieec.uab.es}
\affiliation{Instituci\`{o} Catalana de Recerca i Estudis Avan\c{c}ats (ICREA)
and Institut de Ciencies de l'Espai (IEEC-CSIC),
Campus UAB, Facultat de Ciencies, Torre C5-Par-2a pl, E-08193 Bellaterra
(Barcelona), Spain and Yukawa Institute, Kyoto University, Kyoto, 606-8502 Japan}

\begin{abstract}
We consider modified $f(R)$ gravity which may unify $R^m$ early-time inflation
with late-time $\Lambda$CDM epoch. It is shown that such model passes the
local tests (Newton law, stability of Earth-like gravitational solution,
very heavy mass for additional scalar degree of freedom) and suggests the
realistic alternative for General Relativity. Various scenarios for future
evolution of $f(R)$ $\Lambda$CDM era are discussed.

\end{abstract}

\pacs{11.25.-w, 95.36.+x, 98.80.-k}

\maketitle

\section{Introduction}

Modified gravity is considered as a very interesting alternative proposal
for dark energy. The attractive property of gravitational dark energy
(for a review, see \cite{review}) is the fact that one should not
introduce some strange matter with negative pressure to describe the
late-time cosmic acceleration. The (sudden) change of decelerated
expansion to accelereted one is explained by the change of the properties
of a gravitational theory in the course of the universe evolution. In
other
words, some sub-leading gravitational action terms \cite{CDTT,NO,review}
may become essential ones at the late universe. That is why there is much
activity in the study of different versions of modified $f(R)$ gravity
with applications to dark energy
cosmology \cite{review,CDTT,NO,ANO,FR,FR1,cap,lea}.

Recently, a very realistic modified $f(R)$ gravity which evades the
solar system tests was proposed in ref.\cite{HS} (for related discussion,
see \cite{AB}). On the same time, such
theory leads to an effective $\Lambda$CDM epoch which complies with
observational data with the same accuracy as the usual General Relativity (GR)
with cosmological constant. Generalization of the model \cite{HS} done in
ref.\cite{Uf} suggests a quite natural unified description of early-time
inflation with late-time acceleration following the earlier proposal of
ref.\cite{NO}. In the present work we propose another class of modified
$f(R)$ gravity which unifies $R^m$ inflation with $\Lambda$CDM era. It is
shown that such theory may pass the local tests. The cosmological
properties of such model are studied. Some speculative remarks about the
possibility of a (future/past) anti-gravity phase are made.
It is also shown that such theory may be more friendly with
future observational
data: unlike to usual GR the current $\Lambda$CDM epoch induced by such
theory may enter to future
quintessence/transient phantom era by the effective
reconstruction of the action. Alternatively, one can have the eternal
$\Lambda$CDM epoch as a result of such reconstruction.

\section{Unifying $R^m$ inflation with $\Lambda$CDM cosmology}

We start from the following action of general $f(R)$ gravity
\be
\label{XXX7}
S=\frac{1}{\kappa^2}\int d^4 x \sqrt{-g} \left(R + f(R)\right)\ .
\ee
The equation of motion in $f(R)$-gravity with matter is given by
\be
\label{XXX22}
\frac{1}{2}g_{\mu\nu} F(R) - R_{\mu\nu} F'(R) - g_{\mu\nu} \Box F'(R)
+ \nabla_\mu \nabla_\nu F'(R) = - \frac{\kappa^2}{2}T_{(m)\mu\nu}\ .
\ee
Here $F(R)=R+f(R)$ and $T_{(m)\mu\nu}$ is the matter energy-momentum tensor.
By introducing the auxilliary field $A$ one may rewrite the action
(\ref{XXX7}) in the following form:
\be
\label{XXX10}
S=\frac{1}{\kappa^2}\int d^4 x \sqrt{-g} \left\{\left(1+f'(A)\right)\left(R-A\right)
+ A + f(A)\right\}\ .
\ee
As is clear from (\ref{XXX10}), if $F'(R)=1+f'(R)<0$,
$\kappa_{\rm eff}^2 \equiv \kappa^2 / F'(A)$ becomes negative and the
theory enters the anti-gravity regime. Note that it is not the case for usual GR.

Recently a viable $f(R)$ model has been proposed in ref.\cite{HS}(for
other recent
proposals/study of properties of viable $f(R)$ gravity,
see \cite{cap,lea,AB}). In this model $f(R)$ is chosen to be
\be
\label{HS1}
f_{HS}(R)=-\frac{m^2 c_1 \left(R/m^2\right)^n}{c_2 \left(R/m^2\right)^n + 1}\ ,
\ee
which satisfies the conditions
\be
\label{HS2}
\lim_{R\to\infty} f_{HS} (R) = \mbox{const}\ ,\quad \lim_{R\to 0} f_{HS}(R) = 0\ .
\ee
The second condition means that there is a flat spacetime solution
(vanishing cosmological constant).
The estimation of ref.\cite{HS} suggests that $R/m^2$ is not so small but rather large even
in the present universe and $R/m^2\sim 41$. Hence,
\be
\label{HSb1}
f_{HS}(R)\sim - \frac{m^2 c_1}{c_2} + \frac{m^2 c_1}{c_2^2} \left(\frac{R}{m^2}\right)^{-n}\ ,
\ee
which gives an ``effective'' cosmological constant $-m^2 c_1/c_2$ and
generates the late-time accelerating expansion.
One can show that
\be
\label{UUU0}
H^2 \sim \frac{m^2 c_1 \kappa^2}{c_2} \sim \left(70 \rm{km/s\cdot pc}\right)^2
\sim \left(10^{-33}{\rm eV}\right)^2\ .
\ee
Hence, the above model describes an effective $\Lambda$CDM cosmology.

Although the model \cite{HS} is very succesful,
 early time inflation is not included there.
We have suggested the modified gravity model to treat the inflation and
the late-time accelerating expansion in a unified way \cite{Uf}.
We have considered a simple extention of the model \cite{HS}
to include the inflation at the early universe.
In order to generate the inflation, one may require
\be
\label{Uf1}
\lim_{R\to\infty} f (R) = - \Lambda_i\ .
\ee
Here $\Lambda_i$ is an effective cosmological constant at the early universe and therefore
it is natural to assume $\Lambda_i \gg \left(10^{-33}{\rm eV}\right)^2$.
For instance, it could be $\Lambda_i\sim 10^{20\sim38}\left( {\rm eV}\right)^2$.
In order that the current cosmic acceleration could be generated,
let us consider that currently $f(R)$ is a small constant, that is,
\be
\label{Uf3}
f(R_0)= - 2\tilde R_0\ ,\quad f'(R_0)\sim 0\ .
\ee
Here $R_0$ is the current curvature $R_0\sim \left(10^{-33}{\rm
eV}\right)^2$.
Note that $R_0> \tilde R$ due to the contribution from  matter. In
fact, if
we can regard $f(R_0)$ as an effective cosmological constant, the effective
Einstein equation gives
\be
\label{UUU1}
R_0=\tilde R_0 - \kappa^2 T_{\rm matter}\ .
\ee
Here $T_{\rm matter}$ is the trace of the matter energy-momentum tensor.
We should note that $f'(R_0)$ need not  vanish exactly. Since we are considering
the time scale of one-ten billion years, we only require
$\left| f'(R_0) \right| \ll \left(10^{-33}\,{\rm eV}\right)^4$.
The last condition corresponding to the second one in (\ref{HS2}) is:
\be
\label{Uf4}
\lim_{R\to 0} f(R) = 0\ .
\ee
In the above class of models, the universe starts from the inflation
driven by the effective cosmological
constant (\ref{Uf1}) at the early stage, where curvature is very
large. As curvature becomes smaller, the effective cosmological constant also becomes smaller.
After that  radiation/matter dominates.
When the density of the radiation and  matter becomes smaller and the curvature
goes to the value $R_0$ (\ref{Uf3}), there appears a small effective
cosmological constant (\ref{Uf3}). Hence, the current cosmic expansion could start.

Eq.(\ref{XXX10}) indicates that there could appear an anti-gravity regime
when $1+f'(A)=1+f'(R)<0$.
If we assume that the anti-gravity does not appear through the known
universe history (even in future),
the condition $f'(R)>-1$, combined with the condition (\ref{Uf4}), gives
\be
\label{UU1}
f(R)> - R\ ,
\ee
which could contradict with (\ref{Uf3}). Since 70$\%$ of the total energy
density of the present universe
could be dark energy, we find $\tilde R_0 \sim 0.7 R_0$, that is
\be
\label{UU1b}
f(R_0) \sim - 1.4 R_0 ,
\ee
which seems to conflict with Eq.(\ref{UU1}). The problem could occur even for the model \cite{HS}.
Note, however, (\ref{UU1}) does not always mean that the anti-gravity regime occurs
at current universe. The condition (\ref{Uf4}) might be a condition in future and therefore
the anti-gravity might appear
in future when the curvature becomes smaller than the present value.
More detailed discussion will be given later.

Since the model corresponding to (\ref{Uf1}) has been investigated
in \cite{Uf}, we now propose a model which satisfies
\be
\label{UU2}
\lim_{R\to\infty} f (R) = \alpha R^m \ ,
\ee
with a positive integer $m$ and a constant $\alpha$.
The condition to avoid the anti-gravity $f'(R)>-1$ tells $\alpha>0$ and
therefore $f(R)$ should be positive
at the early universe. On the other hand, Eq.(\ref{Uf3}) or
(\ref{UU1b}) shows that $f(R)$ is negative at
the present universe. Therefore $f(R)$ should cross zero in the past.

At the early universe, if the scalar curvature is large, the $f(R)$-term
could
behave as (\ref{UU2}) and
would dominate the Einstein-Hilbert term (if $m$ is bigger than 1). Now let
assume that there exists matter with an equation of state parameter $w$.
For spatially flat FRW universe,
\be
\label{FRW}
ds^2 = - dt^2 + a(t)^2 \sum_{i=1,2,3} \left(dx^i\right)^2\ ,
\ee
as shown in \cite{ANO}, the scale factor $a(t)$ behaves as
\be
\label{UUU2}
a(t) \propto t^{h_0}\ ,\quad h_0 \equiv \frac{2m}{3(w+1)}\ .
\ee
Then the effective equation of the state parameter $w_{\rm eff}$, which is defined by
\be
\label{UUU3}
w_{\rm eff}= -1 + \frac{2}{3h_0}\ ,
\ee
can be less than $-1/3$ and the accelerating expansion could occur if $m$ is large enough although
$w_{\rm eff} > -1$. Then the inflation could occur due to the $R^m$ behavior of $f(R)$ in (\ref{UU2}).
The expansion of the universe is quintessence like since $w_{\rm eff}>-1$.
 We need not, however,
a real quintessence field as the inflaton. Note, $w$ can vanish, which
corresponds to dust,
that is, cold dark matter or baryons, or $w$ can be equal to $1/3$, which
corresponds to the radiation.
If $m$ is chosen to be large enough, $h_0$ also becomes large and
therefore $w_{\rm eff}$ goes to $-1$, which coresponds to the cosmological constant.
Hence, for such class of models, at the early stage, the universe starts
from the inflation driven by the
$R^m$-behavior, instead of the effective cosmological constant in (\ref{Uf1}).
Since $w_{\rm eff}>-1$,
the curvature could become smaller as time passes and $f(R)$ could cross zero.
At that stage, $f(R)$ could be
neglected and the radiation/matter could dominate. After that,
$f(R)$ becomes negative and its absolute value
could increase as the curvature becomes smaller. When the density
of the radiation and  matter becomes smaller
due to the expansion of the universe and the curvature goes to the value $R_0$
(\ref{Uf3}), there could appear a small effective cosmological constant
(\ref{Uf3}).
Hence, the current cosmic expansion starts.
Note that another possibility to realize the above scenario is to add
sub-leading curvature terms (dominant over $R$) in such a
way that de Sitter like
inflation occurs. Moreover, the reconstruction of such $f(R)$ gravity at
large curvature may be done so that inflationary stage is instable and
curvature-induced exit occurs. As the last possibility to achieve the exit
from the inflationary era one can add (small) non-local gravity action of
the sort recently proposed in refs.\cite{deser,nl}.

A simplest example satisfying the above conditions is
\be
\label{UU2b}
f(R)= \alpha R^m - \beta R^n\ .
\ee
Here $\alpha$ and $\beta$ are positive constant and $m$ and $n$ are positive intergers
satisfying the condition $m>n$. Since
\be
\label{UUU4}
f'(R) = \alpha m R^{m-1} - \beta n R^{n-1}\ ,
\ee
we find
\be
\label{UUU5}
R_0=\left(\frac{\beta n}{\alpha m}\right)^{1/(m-n)}\ ,\quad
f(R_0)=- \beta \left(1 - \frac{n}{m}\right) \left(\frac{\beta n}{\alpha m}\right)^{n/(m-n)}<0\ .
\ee
Since
\be
\label{UUU6}
f(R_0)=-2\tilde R_0 \sim R_0 \sim \left(10^{-33}\,{\rm eV}\right)^2\ ,
\ee
one gets $\alpha \sim R_0^{1-m}$ and
$\beta \sim R_0^{1-n}$. This shows that $f(R)$ becomes larger than $R$, $f(R) \gg R$,
even in the solar system where $R\sim 10^{-61}\, {\rm eV}^2$, which could be inconsistent.

Another, more realistic proposal is
\be
\label{UU2d}
f(R)= \frac{\alpha R^{m+l} - \beta R^n}{1 + \gamma R^l}\ .
\ee
Here $\alpha$, $\beta$, and $\gamma$ are positive constants and $m$, $n$,
and $l$ are positive integers satisfying the condition $m+l>n$.
(If necessary, to achieve the exit from the inflation
more curvature terms with powers less than $m+l$ may be added to
Eq.(\ref{UU2d}).) For simplicity, we now choose
\be
\label{UU3}
m=l=n\ .
\ee
Then since
\be
\label{UU4}
f'(R) = \frac{n R^{n-1}\left( \alpha \gamma R^{2n} - 2\alpha R^n - \beta \right)}{\left(1+\gamma R^n \right)^2}\ ,
\ee
one gets $R_0$ satisfying (\ref{Uf3}) is given by
\be
\label{UUU7}
R_0=\left\{ \left(\frac{1}{\gamma}\right)\left(1+ \sqrt{ 1 + \frac{\beta\gamma}{\alpha} }\right)\right\}^{1/n}\ ,
\ee
and therefore
\be
\label{UU6}
f(R_0) \sim -2 \tilde R_0 = \frac{\alpha}{\gamma^2}\left( 1 + \frac{\left(1 - \frac{\beta\gamma}{\alpha} \right)
\sqrt{ 1 + \frac{\beta\gamma}{\alpha}}}{2 + \sqrt{ 1 + \frac{\beta\gamma}{\alpha}}} \right) \ .
\ee
As a working hypothesis, we assume $\beta\gamma/\alpha \gg 1$,
which will be justified later. Then we have
\be
\label{UUU01}
R_0 \sim \left(\frac{\beta}{\alpha\gamma}\right)^{1/2n}\ ,\quad
f(R_0)= - 2 \tilde R_0 \sim - \frac{\beta}{\gamma}\ .
\ee
One also assumes
\be
\label{UUU02}
f(R_I) \sim \left(\frac{\alpha}{\gamma}\right) R_I \sim R_I\ ,
\ee
when $R$ is given by the scale of the inflation, $R\sim R_I$, which is
$R_I\sim \left(10^{15}\,{\rm GeV}\right)^2 = \left(10^{24}\,{\rm eV}\right)^2$.
The above conditions (\ref{UUU01}) and (\ref{UUU02}) could be solved as
\be
\label{UU9}
\alpha \sim 2 \tilde R_0 R_0^{-2n}\ ,\quad \beta \sim 4 {\tilde R_0}^2 R_0^{-2n} R_I^{n-1}\ ,\quad
\gamma \sim 2 \tilde R_0 R_0^{-2n} R_I^{n-1}\ .
\ee
Then we find $\beta\gamma/\alpha \sim 4 {\tilde R_0}^2 R_0^{-2n} R_I^{2n-2} \sim 10^{228(n - 1)}$,
which is surely large if $n>1$.

The action (\ref{XXX10}) may be presented in scalar-tensor form.
By using the scale transformation $g_{\mu\nu}\to \e^\sigma g_{\mu\nu}$ with
$\sigma = -\ln\left( 1 + f'(A)\right)$, the Einstein frame action follows \cite{NO}:
\bea
\label{XXX11}
&& S_E = \frac{1}{\kappa^2}\int d^4 x \sqrt{-g} \left( R - \frac{3}{2}g^{\rho\sigma}
\partial_\rho \sigma \partial_\sigma \sigma - V(\sigma)\right)\ , \nn
&& V(\sigma) = \e^\sigma g\left(\e^{-\sigma}\right)
 - \e^{2\sigma} f\left(g\left(\e^{-\sigma}\right)\right)
= \frac{A}{F'(A)} - \frac{F(A)}{F'(A)^2}\ .
\eea
Here $F(R)\equiv R + f(R)$ and $g\left(\e^{-\sigma}\right)$ is given by solving
$\sigma = -\ln\left( 1 + f'(A)\right)=\ln F'(A)$ as $A=g\left(\e^{-\sigma}\right)$.
After the scale transformation $g_{\mu\nu}\to \e^\sigma g_{\mu\nu}$,
there appears a coupling of the scalar field $\sigma$ with the matter.
The mass of $\sigma$ is defined by
\be
\label{MN3}
m_\sigma^2 \equiv \frac{1}{2}\frac{d^2 V(\sigma)}{d\sigma^2}
=\frac{1}{2}\left\{\frac{A}{F'(A)} - \frac{4F(A)}{\left(F'(A)\right)^2} + \frac{1}{F''(A)}\right\}\ .
\ee
Unless $m_\sigma$ could not be large, there appears the large correction to the Newton law.

We now investigate the above correction to the Newton law
in the model (\ref{UU2d}) with (\ref{UU9}).
In the solar system, where $R\sim \left(10^{-61}\, {\rm eV}\right)^2$, or in the air on earth,
where $R\sim \left(10^{-50}\, {\rm eV}\right)^2$, we find
\bea
\label{UU11}
&& F(R) = R + f(R) \sim R - 2 R_0 \sim R\ ,\nn
&& F'(R) = 1 + f'(R) \sim 1 + \frac{n\alpha}{\gamma} R^{n-1}
\sim 1 + n\left(\frac{R}{R_I}\right)^{n-1} \sim 1\ ,\nn
&& F''(R) = f''(R) \sim \frac{n(n-1)}{R_I} \left(\frac{R}{R_I}\right)^{n-2}\ .
\eea
Then in the solar system, the mass $m_\sigma$ of the scalar field $\sigma$ is given by
$m_\sigma^2 \sim 10^{-160 + 109 n}\,{\rm eV}^2$ and in the air on the earth,
$m_\sigma^2 \sim 10^{-144 + 98 n}\,{\rm eV}^2$.
In both cases, the mass $m_\sigma$ is very large if $n\geq 2$ and
the correction to the Newton law is very small.

Thus, it is proposed a viable modified gravity which unifies
curvature-induced $R^m$ inflation with effective Lambda-CDM cosmology.
There is no violation of Newton law in such theory while the known
universe expansion history is reproduced.

\section{$\Lambda$CDM era and its future evolution}

Let us discuss further properties of the proposed modified gravity. There
may exist another type of instability (so-called matter instability) in $f(R)$
gravity \cite{DK,Faraoni}. It is known that it is absent in the model of ref.\cite{NO}.
The instability might occur when the curvature is rather large,
as in the planet, compared with
the average curvature at the universe $R\sim \left(10^{-33}\,{\rm eV}\right)^2$.
By multipling Eq.(\ref{XXX22}) with $g^{\mu\nu}$, one obtains
\be
\label{XXX23}
0 = \Box R + \frac{F^{(3)}(R)}{F^{(2)}(R)}\nabla_\rho R \nabla^\rho R
+ \frac{F'(R) R}{3F^{(2)}(R)} - \frac{2F(R)}{3 F^{(2)}(R)} - \frac{\kappa^2}{6F^{(2)}(R)}T\ .
\ee
Here $T\equiv T_{(m)\rho}^{\ \rho}$ and $F^{(n)}(R) \equiv d^nF(R)/dR^n$.
We consider a perturbation from the following solution of the Einstein gravity:
\be
\label{UUU8}
R=R_b\equiv - \frac{\kappa^2}{2}T>0\ .
\ee
Note that $T$ is negative since $|p|\ll \rho$ in the earth and
$T=-\rho + 3 p \sim -\rho$. Then we assume
\be
\label{UUU9}
R=R_b + R_p\ ,\quad
\left(\left|R_p\right|\ll \left|R_b\right|\right)\ .
\ee
Now one can get
\bea
\label{XXX26}
0 &=& -\partial_t^2 R_p + U(R_b) R_p + {\rm const.}\ , \nn
U(R_b)&\equiv& \left(\frac{F^{(4)}(R_b)}{F^{(2)}(R_b)} - \frac{F^{(3)}(R_b)^2}{F^{(2)}(R_b)^2}\right)
\nabla_\rho R_b \nabla^\rho R_b + \frac{R_b}{3} \nn
&& - \frac{F^{(1)}(R_b) F^{(3)}(R_b) R_b}{3 F^{(2)}(R_b)^2} - \frac{F^{(1)}(R_b)}{3F^{(2)}(R_b)}
+ \frac{2 F(R_b) F^{(3)}(R_b)}{3 F^{(2)}(R_b)^2} - \frac{F^{(3)}(R_b) R_b}{3 F^{(2)}(R_b)^2} \ .
\eea
Then if $U(R_b)$ is positive, $R_p$ becomes exponentially large as a function of $t$:
$R_p\sim \e^{\sqrt{U(R_b)} t}$ and the system becomes unstable.
In the model (\ref{UU2d}) with (\ref{UU9}), if $n\geq 2$
\be
\label{Uf12UU}
U(R_b)\sim - \frac{R_I}{3n(n-1)}\left(\frac{R_b}{R_I}\right)^{-n+1}<0\ .
\ee
Therefore there is no such instability in the model under consideration.

Let us consider what occurs when $f'(R)\to -1$. If $f'(R)<-1$, the theory
enters anti-gravity regime as is seen in
(\ref{XXX10}). In the (effective) FRW equations with flat spatial part,
\be
\label{UU14}
\frac{H^2}{3\kappa_{\rm eff}^2}=\rho\ ,\quad 0=\frac{1}{\kappa_{\rm eff}^2}\left(2\dot H + 3H^2 \right) + p\ ,
\ee
anti-gravity means negative $\kappa_{\rm eff}^2$. When $\kappa_{\rm eff}^2<0$,
there is no solution of the FRW equation (\ref{UU14}), which means that the anti-gravity could not
occur in the FRW universe with flat spacial part.
Now we assume $f'(R)=-1$ when $R=R_A > 0$. When $f'(R)\to -1$ but $f'(R)> -1$,
it follows $\kappa_{\rm eff}^2 \to 0$.
Then from (\ref{UU14}), $H$, $\dot H \to 0$ when $\kappa_{\rm eff}^2 \to 0$,
which seems to contradict
 assumption $R_A>0$, since the scalar curvature $R$ vanishes when
$H=\dot H =0$. This indicates that
the scalar curvature $R$ could not reach $R_A$. Then, anyway,
the anti-gravity could not be realized for the real
universe even in the future.
A possibility of the transition between normal gravity and anti-gravity
might be if $\rho$ and $p$ vanishes
when $\kappa_{\rm eff}$ vanishes, then $H$ and/or $\dot H$ might not vanish.
As usual matter gives positive contribution
to $\rho$, one needs the negative contributions to $\rho$. One
contribution might come from the negative cosmological
constant and another might come from the negative spatial curvature, which
gives a contribution to $\rho$ as $- 1/a^2$.
There could be one more possibility for the transition between normal gravity
and anti-gravity, where $H$ vanishes but
$\dot H$ is finite, and therefore the scalar curvature does not vanish. We
should note that $\rho$ could be positive
in the flat spatial geometry but $p$ can vanish or even can be negative as
for  dark energy. Hence, one can speculate that the pre-inflationary era
may
result from the transition from anti-gravity to usual $f(R)$ gravity at
the point with zero effective Newton
coupling and infinite negative cosmological constant.

It is interesting to investigate if one can distinguish the $\Lambda$CDM
epoch from usual GR and the same epoch which appears in
the present $f(R)$-model. The analog of the first FRW equation is:
\be
\label{UU15}
0 = - \frac{F(R)}{2} + 3\left(H^2 + \dot H \right) F'(R)
 - 18 \left(4H^2 \dot H + H \ddot H \right) F''(R)
+ \kappa^2 \rho_{\rm matter}\ .
\ee
For the constant equation of state matter, it is known that  $\rho_{\rm
matter} = \rho_0 a^{-2/3(w+1)}$.
Proposing (\ref{Uf3}), $f(R)$ can be expanded with respect to $R-R_0$ as
\be
\label{UU17}
f(R) = - 2 \tilde R_0 + \delta f\ ,\quad
\delta f \equiv f_0 \left(R - R_0\right)^2 + {\cal O}\left(\left(R-R_0\right)^3\right)\ .
\ee
Here $f_0$ is a positive constant. If we keep only the first term in (\ref{UU17}) by putting $\delta f=0$, we find
the following solution in (\ref{UU15})
\be
\label{UU18}
a=a_0 \e^{g(t)}\ ,\quad
g(t) = g_0(t) \equiv
\frac{2}{3(w+1)} \ln \left( A \sinh \left(\frac{3(1+w)t}{2l}\right) \right)\ .
\ee
Here $A^2 \equiv \rho_0 a_0^{-3(1+w)}/\tilde R_0$, $l^2 \equiv 3/\tilde R_0$.
The solution (\ref{UU18}) is the same as for  Einstein
gravity with a cosmolgical
constant and matter. One now treats $\delta f$ as a perturbation, which
could be justified in near future or near past.
By putting $g(t) = g_0 (t) + \delta g$ and using (\ref{UU17}), Eq.(\ref{UU15}) gives
\bea
\label{UU20}
0 &=& - 6 H_0 \delta \dot g - \frac{2\kappa^2 \rho_{\rm matter\, 0}}{3(w+1)} \delta g
 - \frac{1}{2}\delta f + 3\left(H_0^2 + \dot H_0\right) \delta f' \nn
&& - 18 \left(4H_0^2 \dot H_0 + H_0 \ddot H_0\right) \delta f''
+ {\cal O}\left(\delta g^2\right)\ .
\eea
Here subindex ``$0$'' expresses a quantity given when $\delta f=0$,
especially
\be
\label{UU21}
H_0 = \dot g_0(t)\ ,\quad \rho_{\rm matter\,0} = \rho_0 a_0^{-3(1+w)} \e^{-3(1+w)g_0(t)}\ .
\ee
The solution of (\ref{UU20}) is
\bea
\label{UU22}
\delta g &=& \e^{- \frac{\kappa^2}{9(w+1)}\int^t dt' \frac{\rho_{\rm matter\,0}(t')}{H_0(t')}}
\int^t \frac{dt'}{6H_0(t')}
\left( - \frac{1}{2}\delta f (t') + \left(3H_0(t')^2
+ \dot H_0(t')\right)\delta f'(t') \right. \nn
&& \left. - 18 \left(H_0(t)^2 \dot H_0(t') + H_0(t') \ddot H_0(t')\right) \delta f'' (t') \right)
\e^{ \frac{\kappa^2}{9(w+1)}\int^{t' } dt'' \frac{\rho_{\rm matter\,0}(t'')}{H_0(t'')}}
+ {\cal O}\left(\delta f^2\right)\ .
\eea
Let the present time $t=t_0$. When $t\sim t_0$, we may assume $H_0$, $\dot H_0$, and $\ddot H_0$ are constants.
Furtheremore one may put $\delta f\propto \left(R-R_0\right)^2 \sim 0$ and
$\delta f' \propto R - R_0 \sim 0$, and
$\delta f = 2 f_0$. Then at  leading order with respect to $t-t_0$,
Eq.(\ref{UU22}) has the following form:
\be
\label{UU23}
\delta g \sim - g_0 \left(t - t_0\right)\ ,\quad g_0 \equiv
\frac{6\left(4H_0 \dot H_0 + \ddot H_0\right) f_0}{H_0^2(1+w)}\ ,
\ee
which gives the correction to the standard $\Lambda$CDM model from
$f(R)$-gravity. Here it is assumed $\delta g=0$ when $t=t_0$.
For the model (\ref{UU2d}) with (\ref{UU3}) and (\ref{UU9}), one finds
\be
\label{UU24}
f_0 = \frac{\alpha n^2 R_0^{2n-2}\left(\gamma R_0^n - 1\right)}{\left( 1 + \gamma R_0^n \right)^2}\ .
\ee
 From (\ref{UU9}), we find $\gamma R_0^n \sim 2 \tilde R_0 R_0^{-n} R_I^{n-1} \gg 1$.
Then (\ref{UU24}) could be approximated by
\be
\label{UUU10}
f_0 \sim \frac{\alpha n^2 R_0^{n-2}}{\gamma}\ .
\ee
Then, the order of $g_0$ may be estimated as
\be
\label{UUU11}
g_0 = {\cal O}\left(R_0^{n - 1/2}/R_I^{n-1}\right) \sim 10^{-114n + 81}\,{\rm eV}\ .
\ee
Especially for $n=2$, we find $g_0 \sim 10^{-147}\,{\rm eV}$.
Since 10 Gigayears correspond to
$\left(10^{-33}\,{\rm eV}\right)^{-1}$, $\delta g$ will be of the order
of unity only $10^{115}$ Gigayears
later. Then the correction does not seem to be observable in the near
future or past.
Of course, the linear approximation used in
(\ref{UU20}) or (\ref{UU23}) may be not enough in the model
(\ref{UU2d}) with (\ref{UU9})
and non-linear terms account may be necessary.

In the search for footprints of non-linear modified gravity at the current
epoch one should not forget that the action of such model may be further
modified in the future so that its good properties survive.
One possibility is the following term in the action:
\be
\label{UU28}
\delta f(R) = - \frac{\eta R^p}{R^{p+q} + \zeta}\ .
\ee
Here $p$ and $q$ are positive integer and $\eta$ and $\zeta$ are positive constants satisfying the
conditions $\zeta \ll R_0^{p+q}$ and $\eta \ll R_0^{q+1}$.
We should note that $\delta f(R)$ satisfies the condition (\ref{Uf4}).
Hence, at the present universe
\be
\label{UUU12}
\left| \delta f(R_0)\right| \sim \eta R_0^{-q} \ll R_0\ ,
\ee
and, therefore, the $\delta f(R)$-term can be neglected. When the
curvature is
larger than $R_0$, $\left| f(R) \right|$
is smaller, which shows that the $\delta f(R)$-term is irrelevant in the
past universe. (Note, however, it may be used to improve the observational
predictions of modified gravity when it is necessary).
If the curvature becomes smaller and satisfies the condition
\be
\label{UUU13}
\zeta^{1/(p+q)} \ll R \ll R_0\ ,
\ee
$\delta R$ behaves as
\be
\label{UUU14}
\delta f(R) \sim - \eta R^{-q}\ .
\ee
If $\eta R^{-q} \gg R_0$, which requires,
\be
\label{UUU15}
R_0^{q+1} \gg \eta \gg R_0 R^q \gg R_0 \zeta^{q/(p+q)}\,
\ee
$\delta f(R)$ could dominate at the future universe.
Using the arguments of ref.\cite{ANO}, if matter is included,
we find the future universe may enter to phantom era:
\be
\label{UUU16}
a(t) \sim \left(t_s - t\right)^{h_0}\ ,\quad h_0 \equiv - \frac{2q}{3(w+1)}\ .
\ee
In the above $a(t)$, there seems to appear the Big Rip singularity at $t=t_s$.
Near the singularity, however, the curvature becomes large and
$\delta f(R)$-term does not dominate. Therefore,
the Big Rip singlarity does not occur and the phantom era is transient
one.
After that the universe enters a quintessence or $\Lambda$CDM epoch.

Our consideration shows that even if the current universe (as predicted by
modified gravity under consideration) is qualitatively/quantatively the
same as the standard$\Lambda$CDM era this may not be true in the near
future.
For instance, the (transient) phantom epoch may emerge without the need to
introduce phantom matter.

\section{Discussion}

In summary, we proposed a modified $f(R)$ gravity which predicts natural
unification of early-time inflation with late-time acceleration. This
theory which is closely related with the models \cite{HS,Uf} passes
the local
tests (Newton law, stability of Earth-like gravitational solution, heavy
mass for additional scalar degree of freedom, etc). The speculative
possibility of past or
future anti-gravity regime is briefly mentioned. The evolution of
the $f(R)$
$\Lambda$CDM epoch is discussed. It is shown that it may get out from
the cosmological constant boundary by future reconstruction of
the gravitational action. As a result, the future universe may enter
a quintessence-like or transient phantom era or it may continue to be
asymptotically de Sitter universe forever.
At the next step, it is necessary to investigate the cosmological
perturbations in the highly non-linear gravity under discussion.
However, this is a quite non-trivial task as the results should be
presented
in a gauge-independent formulation (the simple approximation which is
analogous to the one made in GR with dark fluid does not lead to
realistic predictions due to its gauge dependence).
 As more precise observational
data for cosmological parameters are expected very soon, the further study
of various cosmological predictions of our model (to distinguish it from
GR) are requested. This will
be done elsewhere.

\section*{Acknowledgements}

We would like to thank Misao Sasaki for participartion at the early stage
of this work and very useful comments.
The research by S.N. has been supported in part by the
Ministry of Education, Science, Sports and Culture of Japan under
grant no.18549001
and 21st Century COE Program of Nagoya University
provided by Japan Society for the Promotion of Science (15COEG01).
The research by S.D.O. has been supported in part by the projects
FIS2006-02842, FIS2005-01181 (MEC, Spain), by RFBR grant 06-01-00609
(Russia) and by YITP, Kyoto.

\appendix

\section{Consistency of stellar solution in $f(R)$-gravity}

Let us investigate if $f(R)$-gravity admits a consistent stellar solution,
where vacuum solution matches onto the
stellar-interior solution.

For this purpose, one may rewrite Eq.(\ref{XXX22}) in the following form
\be
\label{sol1}
\frac{1}{2}g_{\mu\nu} R - R_{\mu\nu} - \frac{1}{2} g_{\mu\nu} \Lambda + \frac{\kappa^2}{2}T_{(m)\mu\nu}
= - \frac{1}{2}g_{\mu\nu} \left( f(R) + \Lambda \right) + R_{\mu\nu} f'(R)
+ g_{\mu\nu}\Box f'(R) - \nabla_\mu \nabla_\nu f'(R)\ .
\ee
Here $\Lambda$ is the value of $f(R)$ in the present universe, which corresponds to the effective
cosmological constant: $\Lambda = f(R_0)$, then we have $R_0 \sim \Lambda \sim \left(10^{-33}\,{\rm eV}\right)^2$.
We now like to treat the  r.h.s. part as a perturbation. The last two
derivative terms could be dangerous
when we consider the stellar configuration since there could be a jump in
the value of $R$ on the surface of the star.
One may regard the order of the derivative could be the order of the
inverse of the Compton length of a typical
scale of the system.
Since the most dangerous case  corresponds to particle, one may estimate
the order of the derivative could
be the Compton length of proton: $\partial_\mu \sim m_p \sim 1\,{\rm GeV}\sim 10^9\,{\rm eV}$.
Here $m_p$ is the mass of proton.
We also assume the scalar curvature has the order of $R_e \sim
10^{-47}\,{\rm eV}^2$,
which  corresponds to the curvature inside the earth.

First we consider the $1/R$ model:
\be
\label{sol2}
f(R)= - \frac{\mu^4}{R}\ .
\ee
The order of the  dimensional parameter $\mu$ is $10^{-33}\,{\rm eV}$.
One may estimate
\be
\label{sol3}
\Box f'(R) \sim \nabla_\mu \nabla_\nu f'(R) \sim \frac{m_p^2 \mu^4}{R^2} \sim 10^{-20}\,{\rm eV}^2\ ,
\ee
which is much larger than $\Lambda$ or $R_e$ and therefore
the perturbative expansion breaks.

In case of the HS model (\ref{HS1}), we find
\be
\label{sol5}
\Box f'(R) \sim \nabla_\mu \nabla_\nu f'(R) \sim \frac{m_p^2 \Lambda}{m^2}  \left(\frac{R}{m^2}\right)^{-n-1}
\sim 10^{-3 - 17n}\,{\rm eV}^2\ .
\ee
In (\ref{HS1}),  $R/m^2 \sim 41$.
Eq.(\ref{sol5}) shows that if $n>2$, $\Box f(R)$ or $\nabla_\mu \nabla_\nu
f(R)$ could be much smaller than $R$
and therefore the perturbative expansion is consistent.
For the model (\ref{UU2d}) the qualitative structure
is similar to that of
the HS model. This shows that for the class of models under discussion,
the stellar solution is qualitatively similar to the one in Einstein
gravity.

\end{document}